\documentclass[%
 reprint,
 superscriptaddress,
 amsmath,amssymb,
 aps,bm,
nofootinbib]{revtex4-2}
\usepackage{todonotes}
\usepackage{graphicx}
\usepackage{dcolumn}
\usepackage{bm}
\usepackage{mathtools}
\usepackage{physics}
\usepackage{siunitx}
\usepackage{lipsum}  
\usepackage{newtxtext,newtxmath}

\begin{document}

\preprint{APS/123-QED}

\title{Mixing Skyrmions and Merons in Topological Quasicrystals of Evanescent Optical Field}

\author{H.J. Putley}
\email{henry.putley@kcl.ac.uk}
\affiliation{Department of Physics and London Centre for Nanotechnology, King's College London, Strand, London WC2R 2LS, UK}
\author{B. Davies}
\affiliation{%
 Department of Mathematics, Imperial College London, 180 Queen's Gate, South Kensington, London, SW7 2AZ, UK}
\author{F.J. Rodríguez-Fortuño}
\author{A.Yu.~Bykov}
\author{A.V. Zayats}
\affiliation{Department of Physics and London Centre for Nanotechnology, King's College London, Strand, London WC2R 2LS, UK}

 %


\date{\today}

\begin{abstract}
Photonic skyrmion and meron lattices are structured light fields with topologically protected textures, analogous to magnetic skyrmions and merons. Here, we report the theoretical existence of mixed skyrmion and meron quasicrystals in an evanescent optical field. Topological quasiperiodic tilings of even and odd point group symmetries are demonstrated in both the electric field and spin angular momentum. These quasicrystals contain both skyrmions and merons of Néel-type topology. Interestingly, the quasiperiodic tilings are in agreement with the observations of quasiperiodic arrangements of carbon nanoparticles in water driven by ultrasound, and pave the way towards engineering hybrid topological states of light which may have potential applications in optical manipulation, metrology and information processing. 
\end{abstract}

\maketitle
Optical quasiparticles are salient vectorial textures that are both topologically stable and robust. They are characterized by a topological charge that is preserved both in time and under local perturbations \cite{shen2024optical}. Geometrically, they comprise a continuous and differentiable three-component vector field over a confined domain, and are in fact topological solitons \cite{manton2004topological,sugic2021particle}. In the case of 2D quasiparticles, such as skyrmions and merons, the topological charge (or skyrmion number) that characterises the quasiparticle topology is defined as \cite{gobel2021beyond,mcwilliam2023topological}:
\begin{equation}\label{eq:skyrmion_number}
    Q=\frac{1}{4\pi}\int_A s \dd{A}=\frac{1}{4\pi}\int_A\vb{\hat{e}}\vdot\bigg(\pdv{\vb{\hat{e}}}{x}\cross\pdv{\vb{\hat{e}}}{y}\bigg)\dd{A},
\end{equation}
where $A$ defines the 2D quasiparticle domain and $s$ is the skyrmion number density. Here $\vb{\hat{e}}$ is a real, normalized, three-component unit vector that defines the skyrmionic texture. For optical quasiparticles $\vb{\hat{e}}$ can take many forms, with skyrmions and merons observed in the electric field \cite{gutierrez2021optical,shen2022generation}, in spin angular momentum (SAM) \cite{shi2021spin,lei2021optical}, pseudospin \cite{karnieli2021emulating}, and in the Stokes vector field \cite{lin2021microcavity}. Optical quasiparticles have also been observed in light-matter coupled systems, including in momentum space \cite{guo2020meron} and in liquid-crystal optical microcavities \cite{krol2021observation}. They are important for applications ranging from high-resolution imaging and polarisation synthesis to informatin transport, multidimensional quantum entanglement and engineering optical interactions with magnetic domains and quasiparticles. 

In this Letter, we show that a linear superposition of evanescent fields, demonstrated on the example of surface plasmon polaritons (SPPs), can generate optical quasicrystals composed of mixed skyrmions and merons textures in both the electric field and SAM. Our approach goes beyond the previous studies of optical quasiparticle lattices of four- and six-fold rotational symmetries, characteristic of pure meron and skyrmion topologies, respectively, \cite{tsesses2018optical,lei2021photonic,ghosh2021topological,tang2023achiral} and leverages the general theory for constructing quasicrystals in linear wavefields \cite{cherkaev2021wave}. Our work extends the standard method of constructing periodic quasiparticle lattices beyond the restriction of periodic 2D lattices to point groups of higher symmetries that do not necessarily tessellate to fill two-dimensional space. Notably, these quasicrystals are populated by both skyrmions and merons simultaneously, a phenomenon not previously observed in any periodic quasiparticle lattices. The generated quasicrystalline lattices often feature skyrmions at the high-symmetry points, surrounded by merons.



\begin{figure}
    \centering
    \includegraphics[width=.95\columnwidth]{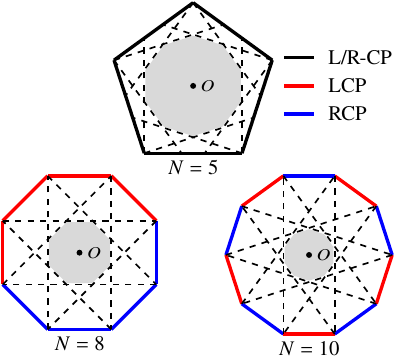}
    \caption{Regular polygons of non-crystallographic point groups of five-fold, eight-fold and ten-fold rotational symmetries. SPPs emanating from the edges interfere in the central region of each polygon (in gray) to form quasicrystals in either $\vb{E}$, $\vb{S}$, or both fields simultaneously. Colors indicate the handedness of chiral gratings for quasicrystals in SAM, where inward-propagating SPPs are excited either by LCP or RCP light.}
    \label{fig:schematic}
\end{figure}

The difficulty in constructing optical quasiparticles is the necessity for the underlying optical field to be non-transverse in nature i.e. in possession of a non-zero longitudinal component \cite{shen2024optical}. Surface-guided modes, including SPPs, fulfill this condition thanks to continuity of the tangential electric and magnetic field components at a metal-dielectric interface, a condition only satisfied by transverse magnetic (TM)-polarized surface waves \cite{plasmonics}. Plasmonic field-skyrmions were the first optical skyrmions to be observed \cite{tsesses2018optical}, and emerged in the form of a field-skyrmion lattice in the evanescent electric field of interfering standing wave SPPs. The standard approach to constructing optical quasiparticle lattices is to model the electric field as the linear superposition of $N$ SPPs incident from the sides of an $N$-sided polygon (Fig.~\ref{fig:schematic}) \cite{tang2023achiral,wang2024graphene}. $N$ also determines the lattice point group, which is linked to the quasiparticle species occupying the unit cell \cite{yu2018transformation,ghosh2021topological,lei2021photonic,zhang2022optical}. Assuming time-harmonic dependence $e^{-i\omega t}$, the electric field at position $\vb{r}=(x,y)$ within the $N$-sided polygon is of the form:
\begin{equation}\label{eq:SPP_waves}
    \vb{E}(\vb{r})=e^{-\lvert k_z\rvert z}\sum_{n=1}^{N} E_n
    \begin{pmatrix}
    i\dfrac{\lvert k_z\rvert}{\lvert k_{\mathrm{SPP}}\rvert}\cos(\frac{2n\pi}{N}) \\[6pt]
    i\dfrac{\lvert k_z\rvert}{\lvert k_{\mathrm{SPP}}\rvert}\sin(\frac{2n\pi}{N}) \\[6pt]
    1
\end{pmatrix} e^{i\phi_n},
\end{equation}
where $k_z=((\omega/c)^2-k^2_{\mathrm{SPP}})^{1/2}$ is the wavevector component normal to the interface, $E_n$ and $\phi_n$ are the amplitude and phase of the $n\mathrm{th}$ propagating SPP, respectively, and 
$ k_\mathrm{SPP}=k_0 \varepsilon_1\varepsilon_2/(\varepsilon_1+\varepsilon_2)^{1/2}$ is the SPP wavenumber with $\varepsilon_1$ being the (positive valued) permittivity of the dielectric occupying $z>0$ and $\varepsilon_2$ being the (negative valued) permittivity of the metal occupying $z<0$. We also assume $\mu=1$ everywhere. In our model, the form of the phase $\phi_n$ depends on our choice of optical field to host quasicrystals, which are non-periodic tilings of the two-dimensional plane that nevertheless exhibit long-range order \cite{penrose1979pentaplexity}. These structures feature symmetries associated with the non-crystallographic point groups, such as the pentagonal or octagonal point groups, which are incompatible with translational symmetries. Quasicrystals can be generated by quasiperiodic functions obtained via the projection method \cite{cherkaev2021wave,de1981algebraic}. This method relies on projecting a periodic function in a higher dimensional space onto a lower dimensional space, such that the resulting pattern maintains the long-range order of the higher-dimensional periodic lattice whilst lacking translational symmetry (see Ref.~\cite{cherkaev2021wave} and its Supplemental Material for details). For our purposes, it is sufficient to assert that the electric field $\vb{E}$ of Eq.~\eqref{eq:SPP_waves} (and/or the SAM $\vb{S}$, depending on our choice of $\phi_n$) is a quasiperiodic function when $N$ corresponds to a non-crystallographic point group \cite{cherkaev2021wave}.

\textit{Electric Field Quasicrystals}---In order to generate quasicrystals in the electric field, the phase $\phi_n$ in Eq.~\eqref{eq:SPP_waves} is modelled as the propagation phase:
\begin{equation}\label{eq:electric_phase}
    \phi_n=-\vb{k}_\mathrm{SPP}^{(n)}\cdot\vb{r},
\end{equation}
where $\vb{k}_\mathrm{SPP}^{(n)}$ is the wavevector of the $n\mathrm{th}$ SPP, and is of magnitude $\lvert k_\mathrm{SPP}\rvert$:
\begin{equation}
    \vb{k}_\mathrm{SPP}^{(n)}=\lvert k_\mathrm{SPP}\rvert\bigg(\cos(\frac{2n\pi}{N}),\sin(\frac{2n\pi}{N})\bigg)^T,
\end{equation}
such that the negative sign in Eq.~\eqref{eq:electric_phase} denotes inward-propagating SPPs. Sources of SPPs carrying a phase of the form of Eq.~\eqref{eq:electric_phase} may constitute diffraction gratings \cite{tsesses2018optical}, surface nanoparticles (such as dipolar sources) \cite{giron2019lateral}, electron beams \cite{gong2014electron} or attenuated total internal reflection coupling devices \cite{plasmonics}. The arrangement of these sources need only correspond to the non-crystallographic point groups, of a kind illustrated in Fig.~\ref{fig:schematic}, for the superposition of SPPs of Eq.~\eqref{eq:SPP_waves} to construct quasiperiodic tilings. Note that for Eq.~\eqref{eq:SPP_waves} to apply, the sources in the simulations must be sufficiently far away from the centre to be in the far field; we take no account of the spatial extent of SPP wavefronts. It should also be noted that by selecting $N=6$ or $N=4$, one reproduces field skyrmions in hexagonal and merons in square lattices, respectively \cite{tsesses2018optical,lei2021photonic}.

\begin{figure*}
    \centering
    \includegraphics[scale=1.1]{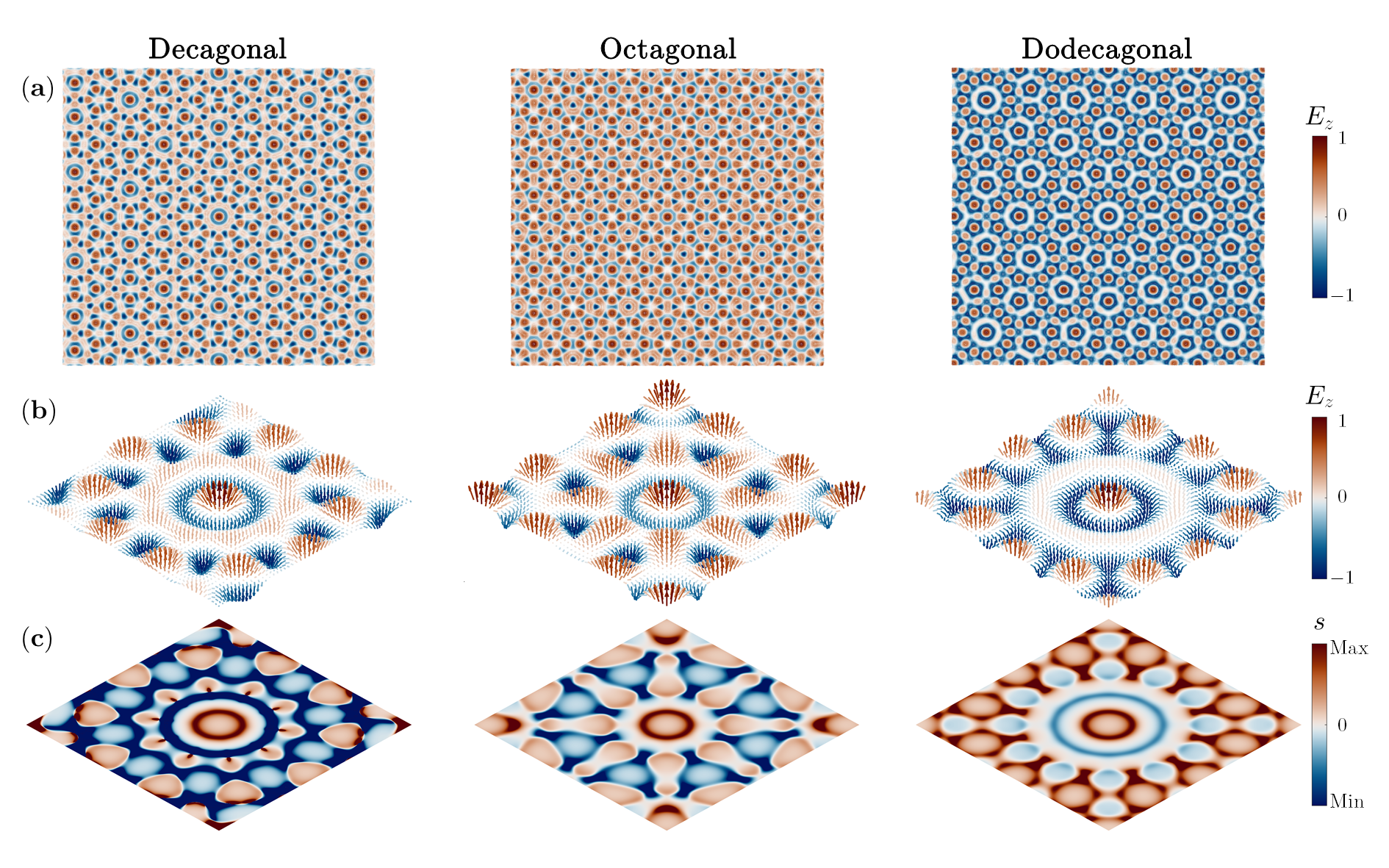}
    \caption{Topological quasicrystals formed by the SPP electric field generated with polygons of $N=5$ (left column), $N=8$ (central column) and $N=12$ (right column) sides: (a) quasicrystalline pattern in $\vb{E}$, (b) the constituent topological quasiparticles at the center of the tiling, and (c) the associated skyrmion number density $s$. Color bars indicate normalized $E_z$ in (a,b) and $s$ in (c). The size of the plots corresponds to $28\lambda_\mathrm{SPP}$.}

    \label{fig:E_QCs}
\end{figure*}

For simplicity, we consider an idealized scenario where SPPs are excited at the interface between air and lossless gold (the presence of losses influences only intensity of the SPP field but not its topology as long as the propagation length exceeds the polygon size). We assume $\varepsilon_1=1$ for air and set the wavelength of light in free-space to be $\lambda=600\,\si{\nano\metre}$, such that the permittivity of lossless gold is $\varepsilon_2=-9.41$ \cite{johnson1972optical} and $\lambda_\mathrm{SPP}=567\,\si{\nano\metre}$. Figure~\ref{fig:E_QCs} depicts quasicrystals in the electric field at the air-gold interface for decagonal, octagonal and dodecagonal symmetries. These quasiperiodic tilings are of a well-known form \cite{penrose1979pentaplexity,cherkaev2021wave} and are rotationally symmetric about the origin of the polygon. In particular, the decagonal tiling we have recovered shares its symmetries with the first Penrose tiling \cite{penrose1979pentaplexity}, and selecting a polygon of $N=10$ produces the same interference pattern as the one shown here for $N=5$. The center of each quasiperiodic tilings can be displaced by altering the amplitudes $E_n$ from $+1$ to $-1$ for different $n$ (not shown here). 

The vectorial field textures in Fig.~\ref{fig:E_QCs}(b) constitute a skyrmion of negative polarity at the centre of each tiling, and an arrangement of merons of alternating polarity surrounding it. The topology of these constituent quasiparticles is fixed throughout the quasicrystal, being the Néel-type texture due to the TM-polarization of the interfering SPPs. To highlight the topological nature of these quasicrystals, 
skyrmion number density $s$ about the centers of the tilings is shown in Fig.~\ref{fig:E_QCs}(c). In this case $s$, and hence $Q$, are defined with respect to the unit electric field $\vb{\hat{e}}=\Re{\vb{E}}/\abs{\vb{E}}$. The skyrmion number density clearly shows distinct domain walls indicative of optical quasiparticles \cite{tsesses2018optical}, and displays the same quasiperiodic symmetries as the electric field. The central skyrmion is of topological charge $Q\approx-1$ and the surrounding merons of topological charge $Q\approx\pm 1/2$, depending on their polarity. These values are approximations due to challenges in defining the domain $A$ for integrating $s$ in Eq.~\eqref{eq:skyrmion_number}, particularly due to the absence of a fundamental cell, and in cases where the cross-sectional area of a quasiparticle is not circular. For the central skyrmion, $A$ was chosen as a circle of radius equalling the ring of $\vb{E}$ pointing along $-z$. For the merons, $A$ was selected by removing all grid points corresponding to a flip in the sign of the $E_z$ component in a local region. The calculated topological charges provide a strong indication of the underlying topological nature of the quasicrystals. The consistency of these values across the tilings supports the observation that they possesses a non-trivial topological character. We also note that in each case, swapping $E_n=\rightarrow-E_n$ for every $n$ inverts the polarity of the quasicrystal and its constituent skyrmions and merons.

\textit{Spin Quasicrystals}---Topological optical lattices in SAM can be generated with chiral plasmonic gratings of ``fishbone nanoapertures'' \cite{tang2023achiral,huang2013helicity} that perform a version of directional SPP excitation \cite{rodriguez2013near,Lin2013polarization}. Under normal illumination by a linearly polarized beam carrying a vortex phase, the chiral grating separates SPPs excited by the left and right circularly polarized components (LCP and RCP) of the beam \cite{tan2020polarization}. The SPPs excited by LCP and RCP light are of the same amplitude but propagate perpendicular to the grating in opposite directions; these directions are determined by the handedness of the chiral grating. 

\begin{figure*}
    \centering
    \includegraphics[scale=1.1]{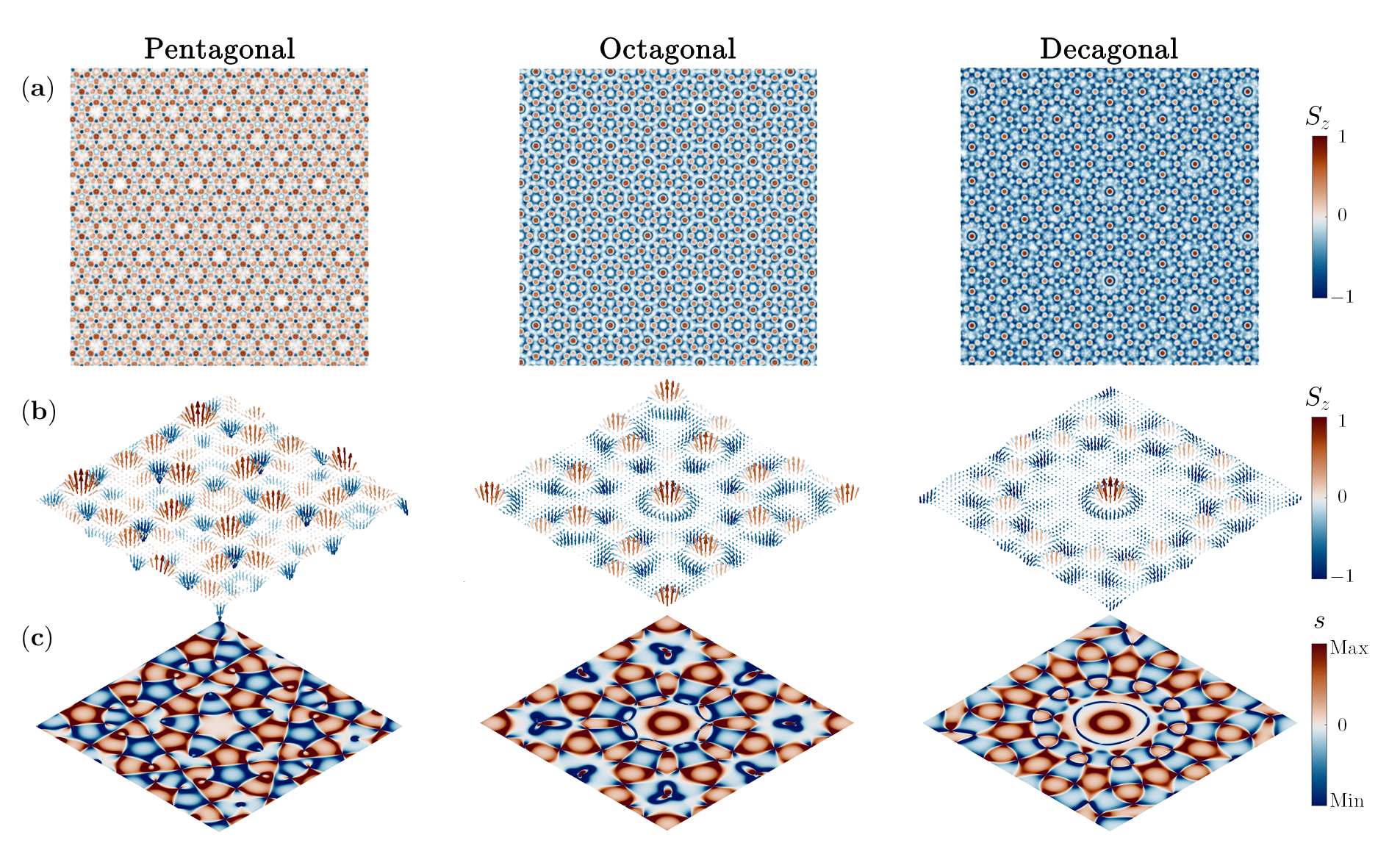}
    \caption{SAM topological quasicrystals generated with polygons of $N=5$ (left column), $N=8$ (central column) and $N=10$ (right column) sides illuminated by a linearly polarized beam of topological charge $l=-1$ and phase $\pi/2$: (a) the quasicrystalline pattern in $\vb{S}$, (b) the constituent topological quasiparticles at high symmetry points, and (c) the associated skyrmion number density $s$. Color bars indicate normalized $S_z$ in (a,b) and $s$ in (c). The size of the plots corresponds to $28\lambda_\mathrm{SPP}$.}
    \label{fig:S_QCs}
\end{figure*}

In the case of quasicrystals, when $N$ is odd as e.g., in the case of pentagon, (Fig.~\ref{fig:schematic}), the handedness of the chiral gratings is uniform with respect to the origin, such that all inward propagating SPPs are excited by the same CP component of the beam. When $N$ is even, the handedness of chiral gratings is altered around the regular polygon (red and blue edges in Fig.~\ref{fig:schematic}) in such a way as to ensure that counter-propagating SPPs are excited by different CP components. The decagon in Fig.~\ref{fig:schematic} can be thought of as the superposition of two regular pentagons of opposite handedness, whereas the octagon is instead two regular polygonal chains of opposite handedness, joined together. Whether $N$ is odd or even, the phase of the $n\mathrm{th}$ SPP is modified with respect to Eq.~\eqref{eq:electric_phase} due to the spin-orbit interaction between the incident beam and the chiral gratings, and is of the form:
\begin{equation}\label{eq:spin_phase}
    \phi_n= -\vb{k}_\mathrm{SPP}^{(n)}\cdot\vb{r} - \sigma_n\bigg(\frac{\pi}{2}+\frac{2n\pi}{N}\bigg) - l\frac{2n\pi}{N}.
\end{equation}
Here, $\sigma_n=+1$ for LCP and $\sigma_n=-1$ for RCP light, $l$ is the topological charge of the incident beam, and $\pi/2$ denotes the polarization of the beam, which points along the $y$-axis. The term in parentheses in Eq.~\eqref{eq:spin_phase} ensures a phase difference of $\pi$ between LCP and RCP excited SPPs, and accounts for rotation of the chiral gratings around the polygon \cite{tan2020polarization,tang2023achiral}.  We have also neglected the contribution of a global phase, as this serves only to rotate the interference pattern. The SAM vector is given by:
\begin{equation}\label{eq:spin}
    \vb{S}=\frac{1}{4\omega}\Im{\varepsilon_0\varepsilon_1(\vb{E}^*\cross\vb{E})+\mu_0(\vb{H}^*\cross\vb{H})}
\end{equation}
where $\vb{H}=-\frac{i}{\omega\mu_0}\curl\vb{E}$. Retaining the material parameters and wavelength of incident light used to generate quasicrystals in the electric field, topological quasicrystals in SAM at the air-gold interface are readily observed (Fig.~\ref{fig:S_QCs}). The pentagonal tiling was generated using only SPPs excited by the RCP component of the beam ($\sigma_n=-1$), and has center of rotation about the origin of the system. This differs to the octagonal and decagonal tilings, which feature a displacement of the spin texture and an off-axis center of rotation. These displacements occur due to the octagonal and decagonal tilings being superpositions of two quasicrystals, each generated by SPPs excited by opposite CP components, whose interference disrupts rotational symmetry about the origin. 

The method of generating spin quasicrystals (that is, whether $N$ is odd or even) also has ramifications for the types of quasiparticles populating the spin texture (Fig.~\ref{fig:S_QCs}b,c). When $N$ is odd, as for the pentagonal tiling, the spin texture contains only merons, of topological charge $Q\approx\pm 1/2$, with no occurrence of skyrmions throughout the quasicrystal. This is to be expected, as it is also the case for periodic spin lattices of merons generated with either LCP or RCP light \cite{tang2023achiral}. When $N$ is even, as in the octagonal and decagonal tilings, the quasicrystals feature both merons and skyrmions simultaneously, with skyrmions occupying the high-symmetry points of a tiling and holding a topological charge $Q\approx-1$. As was the case for the hexagonal lattice of spin skyrmions, each set of chiral gratings of opposite handedness contributes a tiling of spin merons to the central region of the polygon, and it is the interference of these tilings of merons that gives rise to spin skyrmions in the quasicrystal. The topologies of these quasiparticles correspond to the Néel-type texture everywhere, and their calculated topological charge is consistent across the tiling. The quasiperiodic structure is also visible in the skyrmion number density $s$ (Fig.~\ref{fig:S_QCs}c), where now $\vb{\hat{e}}=\Re{\vb{S}}/\abs{\vb{S}}$. Again, $s$ features distinct domain walls between quasiparticles, but in the pentagonal case lacks a central circularly symmetric pattern indicative of a skyrmion. Topological quasicrystals of the spin distributions can be generated using vortex beams with the polarity of the spin texture, and hence the topological charges of the constituent quasiparticles, can be inverted by changing $l\rightarrow-l$ of the excitation beam. 

\begin{figure*}
    \centering
    \includegraphics[scale=1.1]{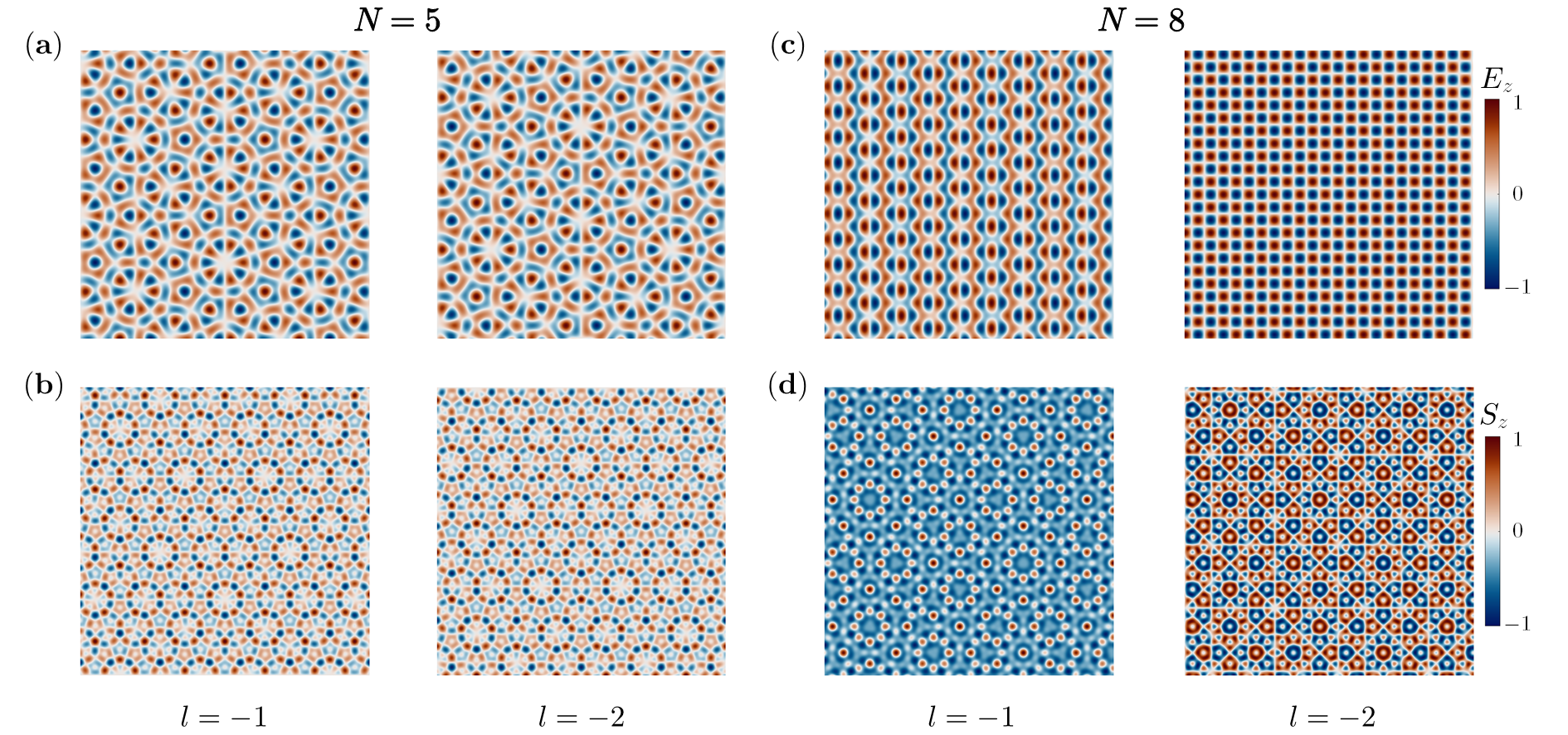}
    \caption{Tilings in $E_z$ (top row) and $S_z$ (bottom row) distributions for polygons of (a,b) $N=5$ sides with $\sigma_n=-1$ for every $n$ and (c,d) $N=8$ sides with $\sigma_n=[+1,+1,+1,+1,-1,-1,-1,-1]$. The topological charge of the excitation beam is labelled by column. The size of the plots corresponds to $28\lambda_\mathrm{SPP}$.}
    \label{fig:odd_even}
\end{figure*}

Lastly, we briefly discuss how modifying the topological charge $l$ of the excitation beam influences the topological ordering of both the electric and spin fields. For quasicrystals of pentagonal symmetry, altering the charge $l$ from $-1$ to $-2$ causes a mirror reflection about the $x$-axis in both the electric and spin fields, and a polarity inversion in the spin field (Figure~\ref{fig:odd_even}a,b). In the case of $N=8$, the tilings constitute periodic lattices in $E_z$, where $l=-1$ generates an oblique lattice of affine basis vectors and $l=-2$ generates a square lattice of $C_2$ rotational symmetry (Figure~\ref{fig:odd_even}c,d). Conversely, the corresponding spin fields reveal two different quasiperiodic tilings that both exhibit octagonal symmetry. The periodicity of the electric field in the case of $N=8$ is due to the interference of two aperiodic tilings, each generated by a polygonal chain of $N=4$ sides (Fig.~\ref{fig:schematic}). The field symmetries of these chains do not correspond to those of a regular polygon, and their interference aligns the quasiperiodic components in such a manner as to bring about a regular lattice with a periodic unit cell. Thus, modification of the topological charge of the incident beam generates diverse tilings of both the electric and spin fields. By adjusting $l$, we have been able to transform the electric field between lattices of oblique and square unit cell, and generate spin fields of different octagonal tilings in the 2D. We also note that similar modifications to the tilings are possible by adjusting the beam polarization (not shown).

\textit{Conclusion}---We have demonstrated that the interference of the evanescent fields of propagating SPPs can generate topological quasicrystals of mixed optical quasiparticle species in both the electric field and SAM. For quasicrystals in the electric field, generated by SPPs carrying only a propagation phase, both skyrmions and merons were present in the vectorial texture, with skyrmions occupying points of high rotational symmetry in the pattern. Topological quasicrystals were generated in SAM by illuminating regular polygons of chiral gratings with a beam carrying intrinsic orbital angular momentum. For odd $N$, this led to quasicrystals populated only by spin merons, whereas for even $N$ the quasicrystals simultaneously featured both spin skyrmions and spin merons. We have also demonstrated control over the texturing of the electric and spin fields by tuning the topological charge of the excitation beam. This led to the observation of periodic lattices in $E_z$ with oblique and square unit cells, and corresponding spin fields of octagonal symmetry. 

While preparing this manuscript for publication, the authors became aware of a preprint discussing quasicrystals in SAM \cite{lin2024photonic}, which detailed the generation of quasiperiodic tilings in $S_z$ by way of the multigrid approach of De Bruijn \cite{de1981algebraic}. The key difference is the demonstration of topological quasiparticles populating both the electric field and spin field quasicrystals in our work. 


H.J.P., F.J.R, A.Yu.B and A.V.Z acknowledge support from by the ERC iCOMM project (789340) and the UKRI EPSRC project EP/Y015673/1. B.D. was supported by the UKRI EPSRC under grant EP/X027422/1.

\bibliography{bib}
\end{document}